\documentclass[12pt,twoside,english]{elsarticle}
\usepackage[T1]{fontenc}
\usepackage[latin9]{luainputenc}
\usepackage{geometry}
\usepackage{amsmath}
\usepackage{amsthm}
\usepackage{amssymb}
\usepackage{stackrel}
\usepackage{graphicx}
\usepackage{setspace}
\makeatletter

\providecommand{\tabularnewline}{\\}

\theoremstyle{plain}
\newtheorem{thm}{\protect\theoremname}
\ifx\proof\undefined
\newenvironment{proof}[1][\protect\proofname]{\par
	\normalfont\topsep6\p@\@plus6\p@\relax
	\trivlist
	\itemindent\parindent
	\item[\hskip\labelsep\scshape #1]\ignorespaces
}{%
	\endtrivlist\@endpefalse
}
\providecommand{\proofname}{Proof}
\fi
\theoremstyle{plain}
\newtheorem{cor}[thm]{\protect\corollaryname}

\journal{Journal A}

\usepackage{algorithm}
\usepackage{algpseudocode}  
\usepackage{lineno}
\usepackage{caption}

\makeatother

\usepackage{babel}
\providecommand{\corollaryname}{Corollary}
\providecommand{\theoremname}{Theorem}

\begin{document}

\begin{frontmatter}{}

\title{Lorentz transformation for the kinematics of degree-4 rigid origami vertices and compatibility of rigid-foldable polygons}

\author[lab1]{Yucai Hu\corref{cor1}}

\ead{huyc@hfut.edu.cn }

\author[lab1]{Licheng Lin}

\author[lab2]{Changjun Zheng}

\author[lab2]{Chuanxing Bi}

\cortext[cor1]{Corresponding author}

\address[lab1]{School of Mechanical Engineering, Hefei University of Technology,
Hefei, Anhui 230009, China}

\address[lab2]{Institute of Sound and Vibration Research, Hefei University of Technology,
Hefei, Anhui 230009, China}
\begin{abstract}
We offer new insight into the folding kinematics of degree-4 rigid origami vertices by drawing an analogy to spacetime in special relativity. 
Specifically, folded states of the vertex, described by pairs of fold angles in terms of cotangent of half-angles, are related through Lorentz transformations in $1+1$ dimensions. 
Linear ordinary differential equations are derived for the tangent vectors on two-dimensional fold-angle planes, with the coefficient matrix depending exclusively on the sector angles. 
By taking the limit to the flat state, we generalize the fold-angle multipliers previously defined for flat-foldable vertices to general and collinear developable degree-4 vertices, and obtain a compatibility theorem on the rigid-foldability of polygons with $n$ developable degree-4 vertices. 
We further explore the rigid-foldable polygons of equimodular type and compose tangent vectors involving fold angles at the creases of the central polygon.
\end{abstract}


\end{frontmatter}{}


\section{Introduction}
Rigid origami permits rotations along the creases while the regions bounded by straight creases, i.e., polygonal facets, undergo no stretch or bend. 
The folding kinematics of rigid origami plays a fundamental role in origami-inspired designs and applications \cite{miura1985method,wei2013geometric,schenk2013geometry,rus2018design,melancon2021multistable}.
As a basis, the rigid folding kinematics of a single vertex has been investigated from several perspectives, such as fold-angle equations using spherical trigonometry \cite{huffman1976,lang2017twists}, spherical linkages \cite{dai1999Mobility,chen2015origami}, quaternions \cite{wu2010modelling},  affine transformations with rotational matrices \cite{belcastro2002modelling}, and deformation map or Lagrangian approach \cite{feng2020designs,farnham2022rigid,grasinger2024lagrangian}, etc. 
The developable degree-4 (DD4) vertices are of particular interest, which possess a single degree of freedom (DOF). 
Using Weierstrass substitutions, Huffman's fold-angle equations for DD4 vertices were further simplified and extended \cite{lang2017twists,fosch2022,hu2023parametric}. 

Given that the entire origami pattern is rigid-foldable, the crease pattern surrounding each facet must also be rigid-foldable. For polygonal facets with DD4 vertices, the pattern is generally non-rigid-foldable due to the overconstraints imposed by the single DOF of each vertex and the loop closure condition \cite{schief2008integrability,tachi2009generalization}.
Hence, it is natural to seek the compatibility conditions enabling rigid-foldable polygons. 
The compatibility condition for the quadrilateral with flat-foldable vertices was first derived by Tachi \cite{tachi2009generalization,tachi2010freeform_quad}.
The result distills to that product of fold-angle multipliers for all flat-foldable vertices equals one \cite{tachi2010freeform_quad,evans2015rigidly,feng2020designs}, which has also been generalized to general polygons with $n$ flat-foldable vertices \cite{evans2015rigidly,evans2015gadgets}. 

On the other hand, rigid-foldable or flexible polyhedra were studied
more than one hundred years ago \cite{Bricard1897}; 
notably, the biquadratic equation for adjacent fold angles of degree-4 vertices which are generally
non-developable was derived by Bricard \cite{Bricard1897}. 
The classification of rigid-foldable Kokosakis polyhedra, which are polyhedral surfaces consisting of one
central planar polygon with degree-4 vertices generally non-developable, has been pursued for many years \citep{sauer1931flachenverbiegung,kokotsakis1933bewegliche,Stachel2010Kokotsakis,nawratil2011reducible},
and was fully achieved by Izmestiev \citep{Izmestiev2017Classification}.
The Kokotsakis polyhedron becomes origami if all the vertices are developable. 
Several approaches have also been proposed to construct larger rigid-foldable quadrilateral patterns, including optimization schemes based on flat-foldable vertices \cite{tachi2010freeform,dudte2016programming,lang2018rigidly,hu2020Constructing,dang2022inverse}, jigsaw puzzle design using symmetry-related DD4 vertices \cite{dieleman2020jigsaw}, and the stitching of rigid-foldable Kokotsakis quadrilaterals from Izmestiev's classification \cite{he2020rigidII}, among others. 
Nevertheless, the rational tiling of rigid-foldable quadrilaterals with different types of DD4 vertices remains poorly understood. 
Furthermore, incorporating general $n$-gons into origami patterns would greatly expand the design space. 
However, the compatibility conditions for rigid-foldable $n$-gons are still unknown, making their tilings into larger patterns even more challenging.

\begin{figure}
	\centering
	\includegraphics[width=0.9\textwidth]{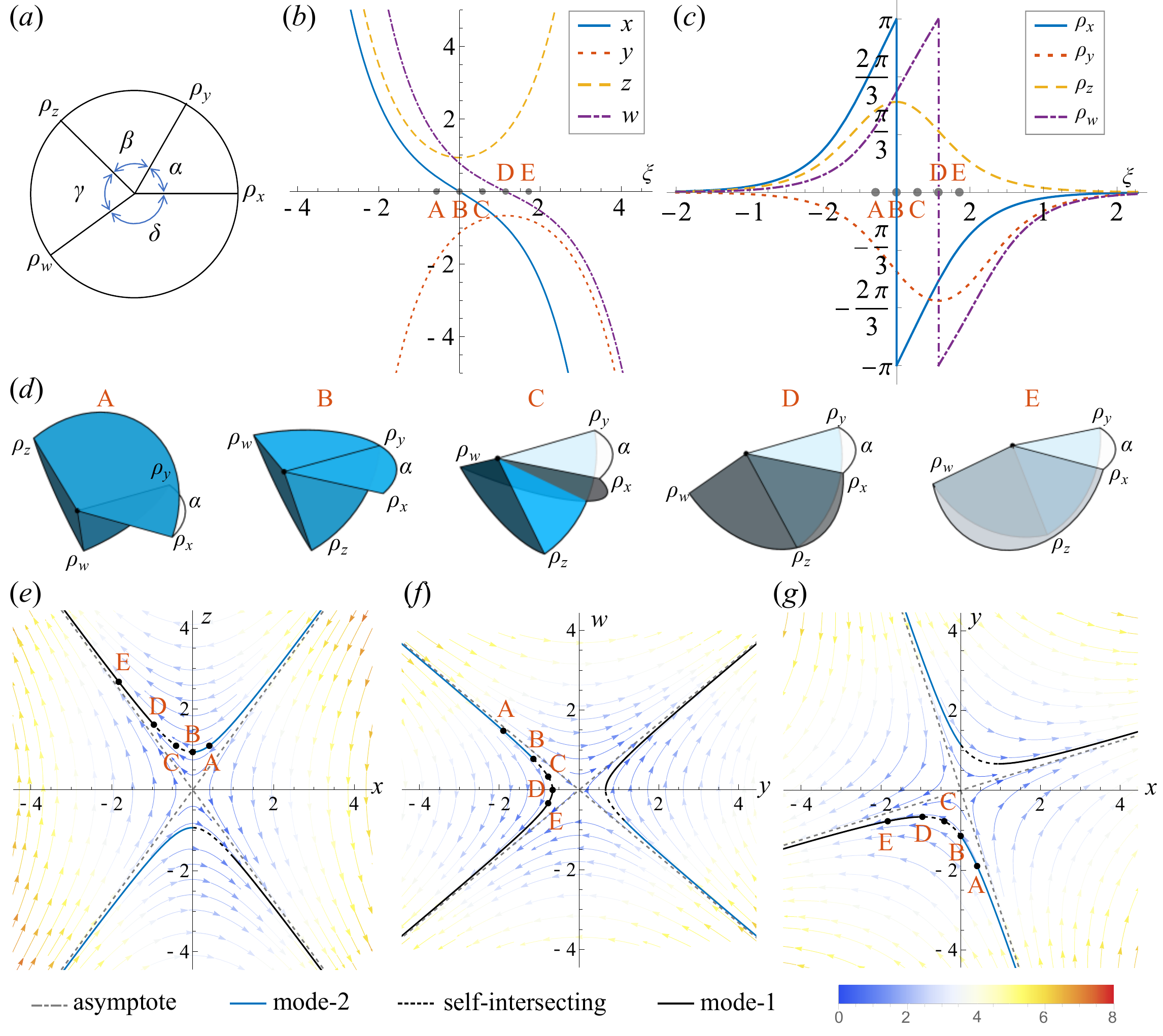}
	\caption{\label{fig:general-vertex}
		Kinematics of a general vertex with sector
		angles $\{\alpha,\beta,\gamma,\delta\}=\{\pi/3,5\pi/12,9\pi/20,4\pi/5\}$.
		(a) shows the crease pattern. (b) and (c) plot the fold angles in
		terms of $\cot(\rho_{r}/2)$ and $\rho_{r}$ versus $\xi$, respectively,
		for $r=x,y,z$, and $w$. (d) shows folded forms where the labels
		A through E correspond to those in (b) and (c) at $\xi=\varphi/2\{-1,0,1,2,3\}$
		with $\varphi\approx1.141$. (e), (f), and (g) show the kinematic
		path projected onto the $x$-$z$, $y$-$w$, and $x$-$y$ planes,
		respectively, with the vector field shown in the background.
	}
\end{figure}

In this paper, we provide insight into the rigid-folding kinematics of DD4 vertices by drawing an analogy to the spacetime of special relativity.
Besides, we establish fold-angle multipliers, previously defined for flat-foldable vertices, for general and collinear DD4 vertices, and obtain a compatibility theorem for rigid-foldable polygons with $n$ DD4 vertices. 
We now turn to introduce the compatibility condition for a single DD4 vertex, represented by two sets of algebraic equations \cite{huffman1976,lang2017twists,fosch2022,hu2023parametric}. 
Referring to Fig. \ref{fig:general-vertex}(a,d), the folded states of a DD4 vertex can be specified by the fold angles $\{\rho_{x},\rho_{y},\rho_{z},\rho_{w}\}$, which are the signed deviations from the flat state. 
With the substitution $r=\cot(\rho_{r}/2)$ for $r=x,y,z$ and $w$, the first set of equations relates the two pairs of opposite fold angles:
\begin{equation}
	x^{2}-\frac{z^{2}}{p^{2}}=a \quad\text{and}\quad
	y^{2}-\frac{w^{2}}{q^{2}}=b \label{eq:eop}
\end{equation}
where
\[
\begin{array}{ccc}
	& p=\sqrt{\dfrac{\sin\beta\sin\gamma}{\sin\alpha\sin\delta}}, & a=\dfrac{\sin(\alpha+\gamma)\sin(\gamma+\delta)}{\sin\beta\sin\gamma},\\
	\textrm{and} & q=\sqrt{\dfrac{\sin\gamma\sin\delta}{\sin\alpha\sin\beta}}, & b=\dfrac{\sin(\alpha+\gamma)\sin(\beta+\gamma)}{\sin\gamma\sin\delta}.
\end{array}
\]
The second set of equations linearly relates three adjacent fold angles:
\begin{subequations}
	\label{eq:ead}
	\begin{equation}
		x\sin(\beta+\gamma)=w\sin\alpha+y\sin\delta;\label{eq:ead_1}
	\end{equation}
	\begin{equation}
		y\sin(\gamma+\delta)=x\sin\beta+z\sin\alpha;\label{eq:ead_2}
	\end{equation}
	\begin{equation}
		z\sin(\delta+\alpha)=y\sin\gamma+w\sin\beta;\label{eq:ead_3}
	\end{equation}
	\begin{equation}
		w\sin(\alpha+\beta)=z\sin\delta+x\sin\gamma.\label{eq:ead_4}
	\end{equation}
\end{subequations}
Equations in (\ref{eq:eop}) are generally hyperbolas and may degenerate into two-intersected lines depending on whether $a$ or $b$ is zero \cite{hu2023parametric}. 
It is known that the folding kinematics of DD4 vertices depend on sector angles, particularly when the sector angles are related. 
Concretely, flat-foldable/collinear vertices, where opposite/adjacent angles supplement to $\pi$, exhibit different folding motions compared to those of general vertices \cite{hu2023parametric, waitukaitis2015origami, Waitukaitis2016}. 
For non-self-intersecting folded states of general vertices, one fold angle (called the unique fold) has an opposite sign to the other three fold angles; 
the unique fold and its opposite fold are termed the major folds, while the other two folds with the same sign are termed the minor folds \cite{huffman1976,Waitukaitis2016,lang2017twists}.
Besides, general DD4 vertices have two folding modes: mode-1, where crease-($x,z$) are the major folds, and mode-2, where crease-($y,z$) are the major folds.
For vertices with collinear creases, trivial folding occurs when the vertex folds along the collinear creases, such folding mode is excluded from the following discussions.

\section{Lorentz transformations and tangent vectors} 
Let $\{\bar{x},\bar{y},\bar{z},\bar{w}\}$ be an initial folded state other than the flat state. 
Since $a$ and $b$ in Eq. (\ref{eq:eop}) depend only on the sector angles and stay constant throughout the folding process, an analogy can be drawn between $a$ and $b$ with spacetime interval in special relativity \cite{schutz2022first}. Hence, we have (see Methods for the derivation)
\begin{thm} \label{thm:LT}
	Consider the nontrivial rigid folding of DD4 vertices.
	Let $\{x,y,z,w\}$ and $\{\bar{x},\bar{y},\bar{z},\bar{w}\}$ be two
	connected folded states on the kinematic path, i.e., they lie on the
	same branch of hyperbola for non-flat-foldable vertices or on the
	same ray emitted from the origin for flat-foldable vertices,
	then pairs of fold angles for the two folded states are related through Lorentz transformations. 
	For the two pairs of opposite fold angles,
	\begin{align}
		&\begin{Bmatrix}x\\z\end{Bmatrix}
		=\begin{bmatrix}\cosh\xi & -\dfrac{1}{p}\sinh\xi\\
			-p\sinh\xi & \cosh\xi
		\end{bmatrix}\begin{Bmatrix}\bar{x}\\
			\bar{z}
		\end{Bmatrix}=\mathbf{L}_{p}\begin{Bmatrix}\bar{x}\\
			\bar{z}
		\end{Bmatrix},\label{eq:LT-r13}\\
		&\begin{Bmatrix}y\\w\end{Bmatrix}
		=\begin{bmatrix}\cosh\xi & \dfrac{1}{q}\sinh\xi\\
			q\sinh\xi & \cosh\xi
		\end{bmatrix}\begin{Bmatrix}\bar{y}\\
			\bar{w}
		\end{Bmatrix}=\mathbf{L}_{q}\begin{Bmatrix}\bar{y}\\
			\bar{w}
		\end{Bmatrix};\label{eq:LT-r24}
	\end{align}
	and for the two adjacent fold angles \textup{$\{x,y\}$},
	\begin{equation}
		\begin{Bmatrix}x\\
			y
		\end{Bmatrix}=\begin{bmatrix}\cosh\xi+h\sinh\xi & u\sinh\xi\\
			v\sinh\xi & \cosh\xi-h\sinh\xi
		\end{bmatrix}\begin{Bmatrix}\bar{x}\\
			\bar{y}
		\end{Bmatrix}=\mathbf{L}_{h}\begin{Bmatrix}\bar{x}\\
			\bar{y}
		\end{Bmatrix}.\label{eq:LT-r12}
	\end{equation}
	with
	\[
	h=\sqrt{\dfrac{\sin\beta\sin\delta}{\sin\alpha\sin\gamma}},u=\dfrac{\sin(\alpha+\beta)}{\sin\beta}h,\;\textrm{and}\;v=\dfrac{\sin(\beta+\gamma)}{\sin\delta}h.
	\]
	In the above expressions, $\mathbf{L}_{p}$, $\mathbf{L}_{q}$, and $\mathbf{L}_{h}$ are self-defined, and $\xi\in\mathbb{R}$ is the free parameter.
\end{thm}

By taking the derivative of $\{x,z\}$ with respect to $\xi$, we can obtain the following linear ordinary differential equations (ODEs) for the tangent vectors on two-dimensional fold-angle planes.
\begin{cor}
	Tangent vectors on the x-z, y-w, and x-y planes can be expressed as
	\begin{align}
		&\begin{Bmatrix}\dfrac{dx}{d\xi}\\
			\dfrac{dz}{d\xi}
		\end{Bmatrix}=\begin{bmatrix}0 & -\dfrac{1}{p}\\
			-p & 0
		\end{bmatrix}\begin{Bmatrix}x\\
			z
		\end{Bmatrix}=\mathbf{X}_{p}\begin{Bmatrix}x\\
			y
		\end{Bmatrix},\label{eq:curve-r13}\\
		&\begin{Bmatrix}\dfrac{dy}{d\xi}\\
			\dfrac{dw}{d\xi}
		\end{Bmatrix}=\begin{bmatrix}0 & \dfrac{1}{q}\\
			q & 0
		\end{bmatrix}\begin{Bmatrix}y\\
			w
		\end{Bmatrix}=\mathbf{X}_{q}\begin{Bmatrix}y\\
			w
		\end{Bmatrix}\label{eq:curve-r24}
	\end{align}
	and
	\begin{equation}
		\begin{Bmatrix}\dfrac{dx}{d\xi}\\
			\dfrac{dy}{d\xi}
		\end{Bmatrix}=\begin{bmatrix}h & u\\
			v & -h
		\end{bmatrix}\begin{Bmatrix}x\\
			y
		\end{Bmatrix}=\mathbf{X}_{h}\begin{Bmatrix}x\\
			y
		\end{Bmatrix}\label{eq:curve-r12}
	\end{equation}
	where $\mathbf{X}_{p}$, $\mathbf{X}_{q}$ and $\mathbf{X}_{h}$ are self-defined. 
	The matrices $\mathbf{L}_{r}$ and $\mathbf{X}_{r}$ for $r=p$, $q$, and $h$ are related through
	\begin{equation}
		\mathbf{X}_{r}=\left.\dfrac{d\mathbf{L}_{r}}{d\xi}\right|_{\xi=0}\;\textrm{and}\;\mathbf{L}_{r}=\exp(\xi\mathbf{X}_{r}).\label{eq:exp-mat}
	\end{equation}
\end{cor}

For general DD4 vertices, a particular choice for the initial state is the state that binds on crease-$z$ or crease-$x$,
\begin{equation}
	\{\bar{x},\bar{z},\bar{y},\bar{w}\}=\begin{cases}
		\varepsilon\sqrt{a}\{1,0,\frac{-\sin\beta}{\sin(\alpha+\beta)},\frac{\sin\gamma}{\sin(\alpha+\beta)}\} & \textrm{if }a>0,\\
		\varepsilon\sqrt{-c}\{0,1,\frac{-\sin\alpha}{\sin(\alpha+\beta)},\frac{\sin\delta}{\sin(\alpha+\beta)}\} & \textrm{if }a<0
	\end{cases}\label{eq:initial-state}
\end{equation}
where 
\[ 
\varepsilon=\pm \quad \text{and} \quad 
c=\frac{\sin(\alpha+\gamma)\sin(\gamma+\delta)}{\sin\delta\sin\alpha}.
\]

By invoking Eqs. (\ref{eq:initial-state}), (\ref{eq:LT-r13}), and (\ref{eq:LT-r24}), the following parametric solution can be obtained for the fold angles of general vertices (see Methods for the derivations, along with the parametric solutions for collinear and flat-foldable vertices).
\begin{cor}
	\label{thm:param-sol-g} For general DD4 vertices, given the initial folded state in Eq. (\ref{eq:initial-state}), the parametric solutions for the fold angles are
	\begin{equation}
		\begin{array}{l}
			x=\varepsilon\sqrt{\left|a\right|}\dfrac{\textup{sign}(a)\exp(\xi)+\exp(-\xi)}{2},\\
			z=-\varepsilon\sqrt{\left|c\right|}\dfrac{\textup{sign}(a)\exp(\xi)-\exp(-\xi)}{2},\\
			y=\textup{sign}(\bar{y})\sqrt{\left|b\right|}\dfrac{\textup{sign}(b)\exp(\xi-\varphi)+\exp(-(\xi-\varphi))}{2},\\
			w=\textup{sign}(\bar{y})\sqrt{\left|d\right|}\dfrac{\textup{sign}(b)\exp(\xi-\varphi)-\exp(-(\xi-\varphi))}{2}
		\end{array}\label{eq:xzyw-sol-exp}
	\end{equation}
	where 
	\[
	d=\dfrac{\sin(\alpha+\gamma)\sin(\beta+\gamma)}{\sin\alpha\sin\beta}
	\quad \textup{and} \quad \varphi=\textup{arctanh}(h^{-\textup{sign}(ab)}).
	\]
\end{cor}
\noindent  
Equation (\ref{eq:xzyw-sol-exp}) indicates that the binding states are located at $\xi=0$ and $\varphi$; thus $\varphi$ is the phase shift between the two binding states. 

As an illustration, we examine the general vertex in Fig.\ref{fig:general-vertex}.
Using Eq.(\ref{eq:xzyw-sol-exp}) with $\varepsilon=+$, Fig.\ref{fig:general-vertex}(b)
and (c) depict the fold angles in terms of $\cot(\rho_{r}/2)$ and
$\rho_{r}$ versus $\xi$, respectively, for $r=x,y,z$, and $w$.
In Fig.\ref{fig:general-vertex}(d), frames A through E are the folded
states at $\xi=\varphi/2\{-1,0,1,2,3\}$ with $\varphi\approx1.141$,
respectively; as shown in frame B and D, the axes of symmetry
for \{$x,z$\} and \{$y,w$\} correspond to the binding states of
the vertex; besides, the two axes are distanced by $\varphi$. 
Fig.\ref{fig:general-vertex}(b), (c), and (d) indicate that the vertex changes from mode-2 to mode-1
as $\xi$ varies from $-\infty$ to $+\infty$. 
Fig.\ref{fig:general-vertex}(e), (f), and (g) show the kinematic path and vector field on the $x$-$z$, $y$-$w$,
and $x$-$y$ planes, respectively. 
The branch with labels is depicted using Eq. (\ref{eq:xzyw-sol-exp}) with $\varepsilon=+$, 
while the remaining branch is depicted using $\varepsilon=-$.
The two branches are not connected in that a state on one branch cannot be transformed into a state on the other branch using the pertinent Lorentz transformation.
The arrows of the streamlines indicate the direction of increasing parameter, and the streamlines are colored
by the magnitude of the vector field, i.e., $\sqrt{(dx/d\xi)^{2}+(dz/d\xi)^{2}}$.
The kinematic path coincides with the streamlines passing through the
binding points \{$x,z$\}=\{$0,$$\pm\sqrt{-c}$\} with $c\approx-0.874$.
For the solid blue segment of the kinematic path, the vertex is of mode-2, as illustrated in frame A in Fig.\ref{fig:general-vertex}(d);
for the solid black segment, the vertex is of mode-1, as illustrated
in frame E; 
the states between B and D are self-intersecting, as illustrated
in frame C. 
The geometrical interpretation of Theorem (\ref{thm:LT}) is that points on the same branch can be transformed into one another using Lorentz transformations. 

\section{Fold-angle multipliers and rigid-foldable polygons} 
Equation (\ref{eq:xzyw-sol-exp}) leads to
\begin{equation}
	\dfrac{z}{x} =-p\dfrac{\exp(\xi)-\textrm{sign}(a)\exp(-\xi)}{\exp(\xi)+\textrm{sign}(a)\exp(-\xi)}=-p(\tanh\xi)^{\textrm{sign}(a)} \label{eq:ratio-zx}
\end{equation}
which indicates that $\tanh\xi$ is proportional to the ratio of $z/x$
or $x/z$ depending on sign($a$). 
Hence, the parameter $\xi$ resembles the rapidity in special relativity \cite{schutz2022first}. 
From Eqs. (\ref{eq:ead_2}) and (\ref{eq:ratio-zx}), the ratio between $y$ and $x$ is
\begin{equation}
	\dfrac{y}{x}=\dfrac{\sin\beta}{\sin(\alpha+\beta)}\left[-\dfrac{\sin\alpha}{\sin\beta}\dfrac{z}{x}-1\right]=\dfrac{\sin\beta}{\sin(\alpha+\beta)}\left[\dfrac{1}{h}(\tanh\xi)^{\textrm{sign}(a)}-1\right]\label{eq:ratio-yx}
\end{equation}
Eqs. (\ref{eq:curve-r12}) and (\ref{eq:ratio-yx}) yield
\begin{equation}
	\dfrac{dy}{dx}=
	\dfrac{\sin\beta}{\sin(\alpha+\beta)}\left[\dfrac{1}{h}(\tanh\xi)^{-\textrm{sign}(a)}-1\right]\label{eq:dydx}
\end{equation}
As the vertex approaches the flat state from mode-1 ($\xi\rightarrow+\infty$) or from mode-2 ($\xi\to-\infty$),
there is
\begin{equation}
	\mu^{\pm}=\lim_{\xi\to\pm\infty}\frac{y}{x}=\dfrac{\sin\beta}{\sin(\alpha+\beta)}(\pm\dfrac{1}{h}-1)=\lim_{\xi\to\pm\infty}\dfrac{dy}{dx}.\label{eq:fa-ratio}
\end{equation}
The ratio in Eq. (\ref{eq:fa-ratio}) can also be written as
\begin{equation}
	\mu^{\lambda}=\tau^{\lambda}\sqrt{\left|\frac{b}{a}\right|}\exp(-\lambda\varphi)\label{eq:fa-ratio-exp}
\end{equation}
with $\lambda=\pm$ and
\[
\tau^{\lambda} =\frac{\lambda(\textrm{sign}(ab)-1)+(1+\textrm{sign}(ab))}{2}\textrm{sign}(\sin(\gamma+\delta))
	=\begin{cases}
		\textrm{sign}(\sin(\gamma+\delta)) & \textrm{if }\lambda=-;\\
		\textrm{sign}(\sin(\beta+\gamma)) & \textrm{if }\lambda=+.
	\end{cases}
\]
The ratio in Eq. (\ref{eq:fa-ratio}) reduces
to $\mu^{\pm}=\dfrac{\pm\sin\alpha-\sin\beta}{\sin(\alpha+\beta)}$, i.e., the fold-angle multipliers previously defined for flat-foldable vertices which hold for the whole folding process. 
For vertices with crease-$x$ and -$z$ being collinear, mode-2 is forbidden whilst $\mu^{+}$ should be derived from Eq. (\ref{eq:ead_1}). 
Similarly, mode-1 is forbidden for vertices with crease-$y$ and -$w$ being collinear. 
The instantaneous fold-angle multipliers with $\mu^{+}$($\mu^{-}$) from mode-1(mode-2) to the flat state are summarized as
\begin{equation}
		\mu^{\lambda}\in\mathcal{M}(\alpha,\beta,\gamma):=
		\begin{cases}
			\mu^{+}=\dfrac{\sin(\beta+\gamma)}{2\sin\gamma} & \textrm{if }\alpha+\beta=\pi\;\textrm{and}\;\beta+\gamma\neq\pi;\\
			\mu^{-}=\dfrac{-2\sin\beta}{\sin(\alpha+\beta)} & \textrm{if }\alpha+\beta\neq\pi\;\textrm{and}\;\beta+\gamma=\pi;\\
			\mu^{\lambda}=\dfrac{\lambda\sin\alpha-\sin\beta}{\sin(\alpha+\beta)} & \begin{array}{c}
				\textrm{if }\alpha+\gamma=\pi,\;\alpha+\beta\neq\pi,
				\textrm{and}\;\beta+\gamma\neq\pi;
			\end{array}\\
			\mu^{\lambda}=\tau^{\lambda}\sqrt{\left|\frac{b}{a}\right|}e^{-\lambda\varphi} & \begin{array}{c}
				\textrm{if }\alpha+\gamma\neq\pi,\;\alpha+\beta\neq\pi,
				\textrm{and}\;\beta+\gamma\neq\pi.
			\end{array}
		\end{cases}\label{eq:mu}
\end{equation}

\begin{figure}
	\centering
	\includegraphics[width=0.9\textwidth]{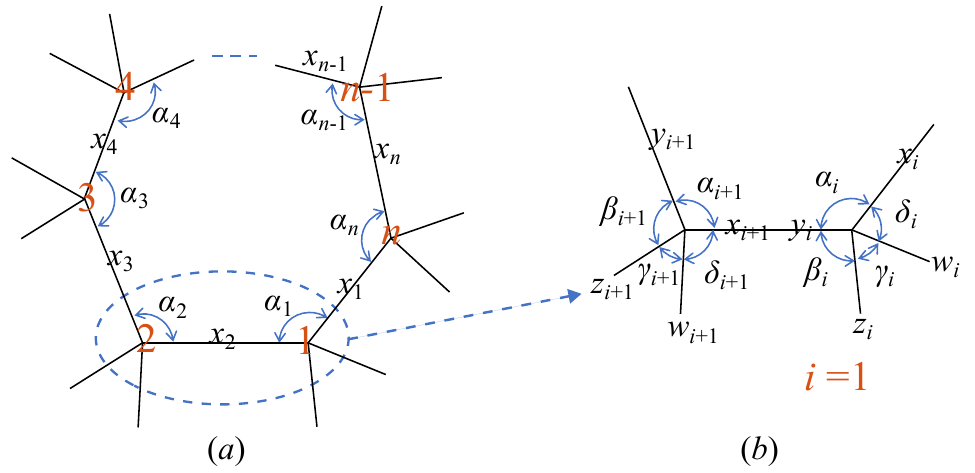}
	\caption{\label{fig:polyhedron-N}Polygon with $n$ DD4 vertices.
		(a) shows that vertices of the $n$-gon where $\alpha_{i}$ is the interior sector angle at vertex-$i$ of the $n$-gon. 
		(b) shows that the four sector angles of vertex-$i$ are labeled counterclockwise as $\alpha_{i},\beta_{i},\gamma_{i}$
		and $\delta_{i}$. Adjacent vertex-$i$ and -($i+1$)
		share the same crease-$y_{i}$ (or -$x_{i+1}$), with the subscript $i+1$ interpreted cyclically.}
\end{figure}

Consider the polygon with $n$ DD4 vertices, as shown in Fig. \ref{fig:polyhedron-N}. 
If the polygon is rigid-foldable, then there exists
$x_{i+1}(t)=y_{i}(t)$ for $i=1,2,\cdots,n$ with $t$ denoting the
folding parameter of the polygon that varies continuously within a region, and
\begin{equation}
	\dfrac{y_{1}}{x_{1}}\dfrac{y_{2}}{x_{2}}\cdots\dfrac{y_{n}}{x_{n}}=1\;\;\textrm{and}\;\;\dfrac{dy_{1}}{dx_{1}}\dfrac{dy_{2}}{dx_{2}}\cdots\dfrac{dy_{n}}{dx_{n}}=1.\label{eq:compatibility}
\end{equation}
Let $\xi_{i}$ be the parameter for vertex-$i$ which follows the
forms of the parametric solutions for general, collinear and flat-foldable vertices (see Eqs. (\ref{eq:xz-sol-g}), (\ref{eq:yw-sol-g}), and (\ref{eq:sol-flatf}) in Methods).
As the polygon folds approach the flat state, $\xi_{i}$ will approach
either $+\infty$ or $-\infty$, depending on the folding mode of vertex-$i$. 
Then, Eq. (\ref{eq:compatibility}) leads to the following result.
\begin{thm}
	\label{thm:fam}If the polygon with $n$ DD4 vertices
	is rigid-foldable from the flat state, then
	\begin{equation}
		\mu_{1}^{\lambda_{1}}\mu_{2}^{\lambda_{2}}\cdots\mu_{n}^{\lambda_{n}}=1
	\end{equation}
	where $\mu_{i}^{\lambda_{i}}\in\mathcal{M}(\alpha,\beta,\gamma)$
	is the instantaneous fold-angle multiplier of vertex-i at the flat
	state, and $\lambda_{i}=+$ (or $-$) when vertex-i is of mode-1 (or
	mode-2). If all n DD4 vertices are general, then
	\begin{subequations}
		\begin{equation}
			\tau_{1}^{\lambda_{1}}\cdots\tau_{n}^{\lambda_{n}}=1\label{eq:fam-gv-tau}
		\end{equation}
		\begin{equation}
			\textrm{and}\quad\sqrt{\left|\frac{b_{1}}{a_{1}}...\frac{b_{n}}{a_{n}}\right|}\exp(-(\lambda_{1}\varphi_{1}+...+\lambda_{n}\varphi_{n}))=1.\label{eq:fam-gv-mag}
		\end{equation}
	\end{subequations}
\end{thm}

\section{Rigid-foldable polygon of equimodular type} 
A polygon of equimodular type requires that the fold angles at the common crease-$y_{i}$ (or -$x_{i+1}$) from the two adjacent general vertices are of the same amplitude \cite{Izmestiev2017Classification}, i.e., 
\begin{equation}
	a_{i+1}=b_{i}\neq0\quad\textrm{or}\quad p_{i+1}=q_{i}\neq1,\label{eq:equi-mag}
\end{equation}
If the polygon is rigid-foldable from the flat state, then Eq. (\ref{eq:fam-gv-mag}) yields
\begin{equation}
	\lambda_{1}\varphi_{1}+...+\lambda_{n}\varphi_{n}=0.\label{eq:phase-shift-0}
\end{equation}
In the context of developable vertices, Eqs. (\ref{eq:phase-shift-0})
and (\ref{eq:fam-gv-tau}) generalize Izmestiev's condition for the
flexible Kokotsakis polyhedron with quadrangular base of equimodular type (see p.734 of \cite{Izmestiev2017Classification}) to $n$-gon base. 
Under the equimodular type, the relationships between $\xi_{i}$ and $t$ can be explicitly expressed. 
Let $\xi_{1}=\sigma_{1}t$,
Eq. (\ref{eq:xzyw-sol-exp}) and $x_{i+1}=y_{i}$ yields
\begin{equation}
	\xi_{i}=\sigma_{i}(t-\sum_{j=1}^{i-1}\sigma_{j}\varphi_{j}).\label{eq:xi_i}
\end{equation}
Eqs. (\ref{eq:xi_i}) and (\ref{eq:xzyw-sol-exp}) indicate that vertex-$i$ binds at $t=\sum_{j=1}^{i-1}\sigma_{j}\varphi_{j}$ and $t=\sum_{j=1}^{i}\sigma_{j}\varphi_{j}$. 
When $t\to+\infty$, $\xi_{i}\to\sigma_{i}\infty$
with $\sigma_{i}=+$/$-$ denoting mode-1/mode-2, and $\lambda_{i}=\textrm{sign}(\xi_{i})=\sigma_{i}$.
With Eq. (\ref{eq:xi_i}), the ratio in Eq. (\ref{eq:ratio-yx}) for vertex-$i$ is
\begin{equation}
	\mu_{i}=\dfrac{y_{i}}{x_{i}}=\tau_{i}\sqrt{\left|\frac{b_{i}}{a_{i}}\right|}\frac{\textrm{sign}(b_{i})\exp(\xi_{i+1})+\exp(-\xi_{i+1})}{\textrm{sign}(a_{i})\exp(\xi_{i})+\exp(-\xi_{i})}
\end{equation}
where
\[
\tau_{i}=\frac{\textrm{sign}(b_{i})(1-\sigma_{i}\sigma_{i+1})+(1+\sigma_{i}\sigma_{i+1})}{2}\textrm{sign}(\sin(\gamma_{i}+\delta_{i})).
\]
Thus, Eq. (\ref{eq:compatibility}) leads to
\begin{equation}
	\tau_{1}\cdots\tau_{n}=1.\label{eq:tau}
\end{equation}
With $f_{i}=\dfrac{\textrm{sign}(a_{i})(1+\sigma_{i})+(1-\sigma_{i})}{2}$, it can be shown that $\tau_{i}^{\lambda_{i}}=\tau_{i}f_{i}f_{i+1}$ when $t\to+\infty$, hence Eq. (\ref{eq:tau}) is identical to Eq. (\ref{eq:fam-gv-tau}) under the equimodular type. 
From the above construction, the conditions in Eqs. (\ref{eq:equi-mag}), (\ref{eq:fam-gv-tau}) (or (\ref{eq:tau})), and (\ref{eq:phase-shift-0}) are sufficient for the equimodular polygon to be rigid-foldable.

\subsection{Composition of tangent vectors}
Additionally, we can successively compose the tangent vectors along creases of the central polygon using Eq. (\ref{eq:curve-r12}), yielding the tangent vector on the $x_1$-$y_{n-1}$ plane as (see Methods for the derivation)
\begin{equation}
	\begin{Bmatrix}\dfrac{dx_{1}}{dt}\\
		\dfrac{dy_{n-1}}{dt}
	\end{Bmatrix}=\begin{bmatrix}h_{1,n-1} & u_{1,n-1}\\
		v_{1,n-1} & -h_{1,n-1}
	\end{bmatrix}\begin{Bmatrix}x_{1}\\
		y_{n-1}
	\end{Bmatrix}=\mathbf{Y}_{1,n-1}\begin{Bmatrix}x_{1}\\
		y_{n-1}
	\end{Bmatrix}\label{eq:dx1dt-dxndt}
\end{equation}
where the entries of $\mathbf{Y}_{1,n-1}$ are
\begin{equation}
	\begin{array}{c}
		\vspace{10pt}
		h_{1,n-1}=(\textrm{tanh}({\sum_{i=1}^{n-1}}\sigma_{i}\varphi_{i})){}^{-\textrm{sign}(a_{1}a_{n})}\\
		\vspace{10pt}
		u_{1,n-1}=\sigma_{1}u_{1}{\prod_{i=2}^{n-1}}\dfrac{-\sigma_{i}u_{i}}{(\tanh({\sum_{j=1}^{i-1}}\sigma_{j}\varphi_{j})){}^{-\textrm{sign}(a_{1}a_{i})}+\sigma_{i}h_{i}};\\
		\vspace{10pt}
		v_{1,n-1}=\sigma_{1}v_{1}{\prod_{i=2}^{n-1}}\dfrac{\sigma_{i}v_{i}}{(\tanh({\sum_{j=1}^{i-1}}\sigma_{j}\varphi_{j})){}^{-\textrm{sign}(a_{1}a_{i})}+\sigma_{i}h_{i}}.
	\end{array}\label{eq:dx1dt-dxndt-entry}
\end{equation}
At vertex-$n$, Eq.(\ref{eq:curve-r12}) holds for \{$x_n,y_n$\}.
If the polygon is rigid-foldable, then $y_{n-1}=x_{n}$ and $x_{1}=y_{n}$, thus
\begin{equation}
	\mathbf{Y}_{1,n-1}=\sigma_{n}\begin{bmatrix}-h_{n} & v_{n}\\
		u_{n} & h_{n}
	\end{bmatrix}.\label{eq:tangent-vector-compat}
\end{equation}

\subsection{Example: rigid-foldable quadrilateral of equimodular type} 
\begin{figure}
	\centering
	\includegraphics[width=0.95\textwidth]{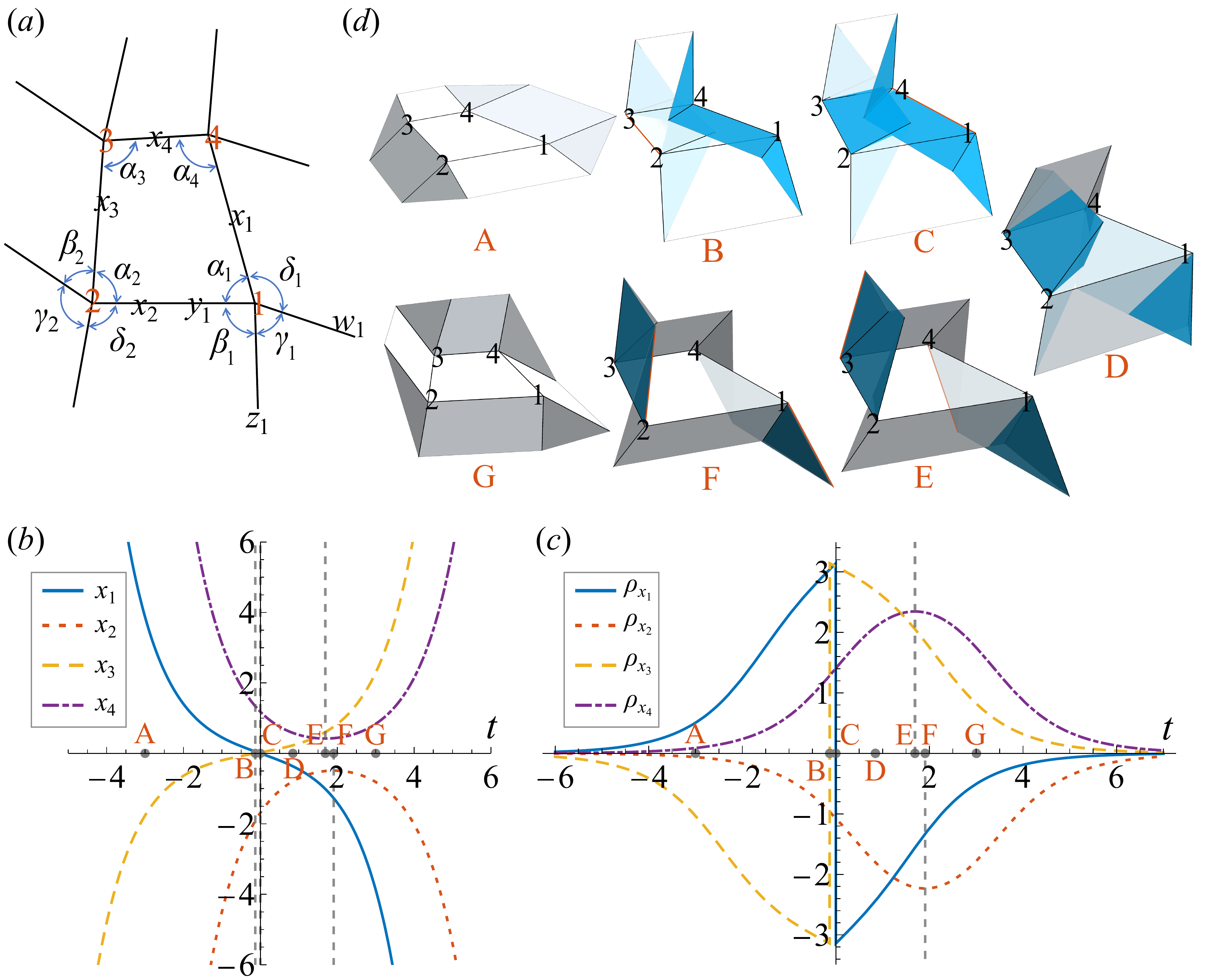}
	\caption{\label{fig:4gon}Rigid-foldable quadrilateral of equimodular type: fold angles and folded forms. 
		(a) shows the crease pattern. 
		(b) plots the fold angles at the four creases of the central quadrilateral,
		i.e., $x_{i}$ for $i=1,2,3$ and 4, and (c) is the counterpart for
		$\rho_{x_{i}}$. 
		(d) shows folded forms at $t=\{-3,\varphi_{1}-\varphi_{2},0,\varphi_{4}/2,\varphi_{4},\varphi_{1},3\}$
		with $\{\varphi_{1},\varphi_{2},\varphi_{4}\}\approx\{1.908,2.039,1.691\}$. 
	}
\end{figure}

\begin{figure}
	\centering
	\includegraphics[width=0.9\textwidth]{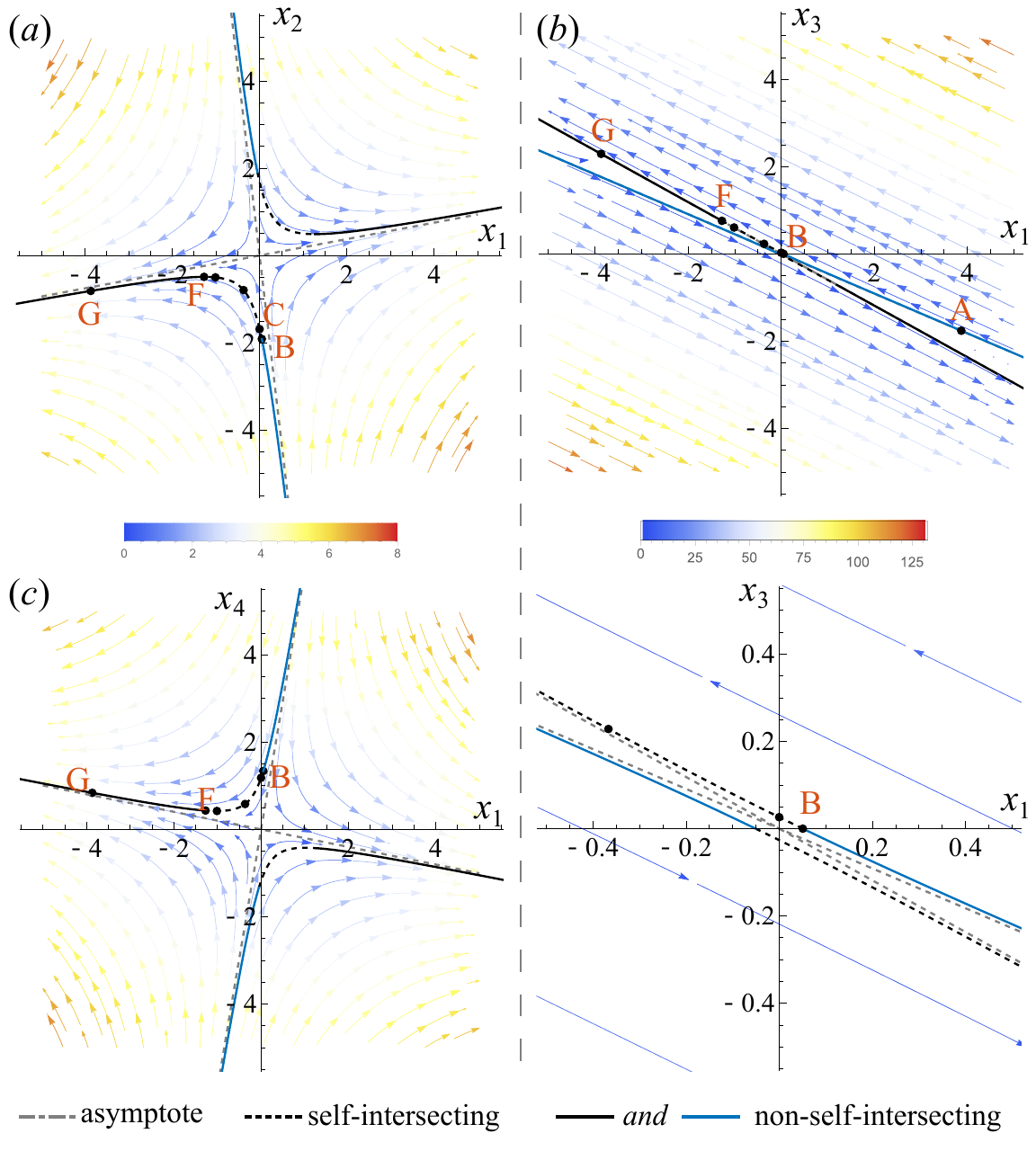}
	\caption{\label{fig:4gon-field} Rigid-foldable quadrilateral of equimodular type: kinematic path and vector field. 
		(a), (b), and (c) plot the results on the $x_{1}$-$x_{2}$, $x_{1}$-$x_{3}$, and $x_{1}$-$x_{4}$ planes, respectively.
		In (b), the bottom subplot magnifies the area near the origin of the upper subplot.
		The streamlines are colored by the magnitude of the vector field.
	}
\end{figure}

Here we examine a rigid-foldable quadrilateral of equimodular type
constructed by using an optimization scheme, see Methods for details on the algorithm. 
Table \ref{tab:sector-angles-k4} lists the sector angles with 16-digit precision such that Eqs.(\ref{eq:equi-mag}),
(\ref{eq:fam-gv-tau}), and (\ref{eq:phase-shift-0}) are satisfied to a precision of $10^{-13}$. 
The crease pattern is shown in Fig.\ref{fig:4gon}(a). 
The fold angles at the creases of the central quadrilateral are plotted in Fig.\ref{fig:4gon}(b) and (c) in terms of $x_{i}=\cot(\rho_{x_{i}}/2)$ and $\rho_{x_{i}}$, respectively, for $i=1,2,3$, and 4. 
Fig.(d) shows several reprentative folded forms at $t=\{-3,\varphi_{1}-\varphi_{2},0,\varphi_{4}/2,\varphi_{4},\varphi_{1},3\}$
with $\{\varphi_{1},\varphi_{2},\varphi_{4}\}\approx\{1.908,2.039,1.691\}$.
For states between B and F ($\varphi_{2}-\varphi_{1}<t<\varphi_{1}$),
at least one of the four vertices self-intersects. 
In frame B, vertices 2 and 3 simultaneously bind at crease-$x_{3}$ with $t=\varphi_{2}-\varphi_{1}$.
Similar scenarios occur in frames C (vertices 4 and 1 bind at crease-$x_{1}$),
E (vertices 3 and 4 bind at creases-$z_{4}$ and -$w_{3}$), and F
(vertices 1 and 2 bind at creases -$w_{1}$ and -$z_{2}$). 
Fig.\ref{fig:4gon-field}(a), (b), and (c) show the kinematic path on the $x_{1}$-$x_{2}$, $x_{1}$-$x_{3}$, and $x_{1}$-$x_{4}$ planes, respectively; the vector field is also
shown in the background using Eq.(\ref{eq:dx1dt-dxndt}). 
The dashed line segment of the kinematic path represents the folded forms in
which at least one vertex is self-intersecting, see Figs.\ref{fig:4gon}(b),
(c), and (d). 
A rigid-foldable pentagon of equimodular type, constructed similarly, is also provided in Methods.
\begin{table}
	\centering
	\caption{\label{tab:sector-angles-k4}Sector angles for the rigid-foldable
		quadrilateral of equimodular type}
	\begin{tabular}{|c|c|c|c|c|}
		\hline 
		Vertex & $\alpha$ & $\beta$ & $\gamma$ & $\delta$\tabularnewline
		\hline 
		\hline 
		1 & 1.3 & 1.6 & 1.217341933495040 & 2.165843373684546\tabularnewline
		\hline 
		2 & 1.5 & $\pi/3$ & 1.987400348664059 & 1.748587407318930\tabularnewline
		\hline 
		3 & 1.7 & 1.289150125833283 & 1.202005510782159 & 2.092029670564144\tabularnewline
		\hline 
		4 & $2\pi-4.5$ & 1.000165606398310 & 1.787849109309900 & 1.711985284291790\tabularnewline
		\hline 
	\end{tabular}
\end{table}

\section{Conclusion} 
This paper reveals that folded states of DD4 vertices are related through Lorentz transformations. 
Based on those transformations, we obtain linear ODEs with constant coefficients, depending solely on the sector angles, for tangent vectors on two-dimensional fold-angle planes, and derive parametric solutions for general, collinear, and flat-foldable vertices.
In addition to flat-foldable vertices, we establish instantaneous fold-angle multipliers at the flat state for general
and collinear vertices and derive a compatibility theorem for the rigid-foldable polygon. 
The compatibility theorem is then used to analyze the rigid-foldable polygons of equimodular type. 

Vertices higher than degree-4 own multi-DOFs, and their rigid-folding kinematics are much more complicated. 
Explicit solutions for the symmetric modes of degree-6 and degree-8 vertices have been recently derived \cite{farnham2022rigid,grasinger2024lagrangian}.
A systematic understanding of DD4 kinematics is helpful since certain modes of higher degree vertices can be reduced to degree-4 \cite{farnham2022rigid}.
Besides, the compatibility theorem on rigid-foldable polygons may inform the design of larger origami patterns.
The similarity between the rigid-folding kinematics of DD4 vertices and spacetime in special relativity also highlights the value of origami as a platform for mathematical and physical education.

\section*{Acknowledgement}
	This work is supported by the National Natural Science Foundation
	of China (Grant Nos. 12102163 and 12272117), the Fundamental Research
	Funds for the Central Universities (Grant No. JZ2024HGTB0236), and
	the Guangdong Basic and Applied Basic Research Foundation (Grant No.
	2021A1515012338).
	
\bibliographystyle{unsrt}
\addcontentsline{toc}{section}{\refname}\bibliography{dd4_lorentz}
\biboptions{sort&compress}

\section*{Methods}
\subsection*{Proof of Theorem \ref{thm:LT}}
Here we prove Theorem \ref{thm:LT}.
For vertices without collinear creases, i.e., $\alpha+\beta\neq\pi$
and $\beta+\gamma\neq\pi$, the transformation between $\{x,z\}$
and $\{\bar{x},\bar{z}\}$, i.e., Eq. (\ref{eq:LT-r13}), can be obtained through the analogy with the Lorentz transformation in $1+1$ dimensions.
Then, invoking  Eqs. (\ref{eq:LT-r13}) and (\ref{eq:ead}), the transformation in Eqs. (\ref{eq:LT-r24}) and (\ref{eq:LT-r12}) can be derived.
Specifically, the relationship between $\{y,w\}$ and $\{x,z\}$, as well as $\{\bar{y},\bar{w}\}$ and $\{\bar{x},\bar{z}\}$, can be obtained from Eqs. (\ref{eq:ead_2}) and (\ref{eq:ead_4}) as
\begin{equation}
	\begin{Bmatrix}y\\w
	\end{Bmatrix}=\dfrac{1}{\sin(\alpha+\beta)}\begin{bmatrix}-\sin\beta & -\sin\alpha\\
		\sin\gamma & \sin\delta
	\end{bmatrix}\begin{Bmatrix}x\\z
	\end{Bmatrix}=\mathbf{T}_{o}\begin{Bmatrix}x\\
		z
	\end{Bmatrix}\label{eq:r2r4-r1r3}
\end{equation}
where $\mathbf{T}_{o}$ is self-defined.
As $\det(\mathbf{T}_{o})=\sin(\beta+\gamma)\csc(\alpha+\beta)$, $\mathbf{T}_{o}$ is nonsingular for vertices without collinear creases. 
From Eqs. (\ref{eq:LT-r13}) and (\ref{eq:r2r4-r1r3}), we have
\begin{equation}
	\begin{Bmatrix}y\\
		w
	\end{Bmatrix}=\mathbf{T}_{o}\mathbf{L}_{p}\begin{Bmatrix}\bar{x}\\
		\bar{z}
	\end{Bmatrix}=\mathbf{T}_{o}\mathbf{L}_{p}\mathbf{T}_{o}^{-1}\begin{Bmatrix}\bar{y}\\
		\bar{w}
	\end{Bmatrix}=\mathbf{L}_{q}\begin{Bmatrix}\bar{y}\\
		\bar{w}
	\end{Bmatrix}.\label{eq:lorentz-2}
\end{equation}
Equation (\ref{eq:LT-r12}) can be derived similarly using the following relationship between $\{x,y\}$ and $\{x,z\}$ which is obtained from Eq. (\ref{eq:ead_2})
\begin{equation}
	\begin{Bmatrix}x\\
		y
	\end{Bmatrix}=\begin{bmatrix}1 & 0\\
		\dfrac{\sin\beta}{\sin(\gamma+\delta)} & \dfrac{\sin\alpha}{\sin(\gamma+\delta)}
	\end{bmatrix}\begin{Bmatrix}x\\
		z
	\end{Bmatrix}=\mathbf{T}_{a}\begin{Bmatrix}x\\
		z
	\end{Bmatrix}.\label{eq:r1r2-r1r3}
\end{equation}
Note that the derivation remains valid for flat-foldable vertices
as long as there are no collinear creases, and in flat-foldable case, $p=q=1$ and $h=\sin\beta \csc\alpha$.

For vertices with crease-$x$ and -$z$ being collinear, i.e., $\alpha+\beta=\pi$
and $\beta+\gamma\neq\pi$, it is clear that $a=0$, $p=h=1$ and $q=\sin\gamma\csc\alpha$. The above derivation is inapplicable since
$\mathbf{T}_{o}$ and $\mathbf{T}_{a}$ become singular. Nevertheless, we can first obtain Eq. (\ref{eq:LT-r24}), and then substitute Eq. (\ref{eq:LT-r24})
into Eq. (\ref{eq:ead_1}), leading to
\begin{equation}
	\begin{aligned}x & =\dfrac{1}{\sin(\beta+\gamma)}\left[(\bar{y}\sin\delta+\bar{w}\sin\alpha)\cosh\xi+(\bar{y}q\sin\alpha+\bar{w}\dfrac{\sin\delta}{q})\sinh\xi\right]\\
		& =\bar{x}\cosh\xi+\dfrac{\bar{y}\sin\gamma+\bar{w}\sin\beta}{\sin(\beta+\gamma)}\sinh\xi\\
		& =\bar{x}\cosh\xi-\bar{z}\sinh\xi.
	\end{aligned}
\end{equation}
Besides, equation (\ref{eq:ead_2}) leads to $z=-x$. 
Hence, Eq. (\ref{eq:LT-r13}) is proven. 
With Eq. (\ref{eq:LT-r24}) and the relationship
\begin{equation}
	\begin{Bmatrix}x\\
		y
	\end{Bmatrix}=\begin{bmatrix}\dfrac{\sin\delta}{\sin(\beta+\gamma)} & \dfrac{\sin\alpha}{\sin(\beta+\gamma)}\\
		1 & 0
	\end{bmatrix}\begin{Bmatrix}y\\
		w
	\end{Bmatrix}=\mathbf{T}_{b}\begin{Bmatrix}y\\
		w
	\end{Bmatrix},
\end{equation}
the transformation on $\{x,y\}$ in Eq. (\ref{eq:LT-r12}) can
be derived similarly as that in Eq. (\ref{eq:lorentz-2}). 
The proof for vertices with crease-$y$ and -$w$ being collinear can be obtained similarly.

\subsection*{Parametric solutions}

This section derives the parametric solutions for the general, collinear, and flat-foldable DD4 vertices. 
For non-flat-foldable vertices with $a>0$, crease-$x$ and -$z$ are not collinear, from Eq. (\ref{eq:LT-r13}),
the fold angle $x$ can be expressed as
\begin{equation}
	\begin{aligned}x & =\sqrt{a}(\dfrac{\bar{x}}{\sqrt{a}}\cosh\xi-\dfrac{\bar{z}}{\sqrt{c}}\sinh\xi)\\
		& =\sqrt{a}(\textrm{sign}(\bar{x})\dfrac{\left|\bar{x}\right|}{\sqrt{a}}\cosh\xi-\dfrac{\bar{z}}{\sqrt{c}}\sinh\xi)\\
		& =\textrm{sign}(\bar{x})\sqrt{a}\cosh(\xi-\bar{\varphi}_{x})
	\end{aligned}
\end{equation}
where $\tanh\bar{\varphi}_{x}=\dfrac{1}{p}\dfrac{\bar{z}}{\bar{x}}$.
For non-flat-foldable vertices with crease-$x$ and -$z$ being collinear,
i.e., $\alpha+\gamma\neq\pi$, $\alpha+\beta=\pi$, and $\beta+\gamma\neq\pi$,
we have $a=0$, $b\neq0$, and $p=1$. Eq. (\ref{eq:ead_1}) leads
$\bar{z}=-\bar{x}$, hence
\begin{equation}
	x=\bar{x}\cosh\xi-\bar{z}\sinh\xi=\bar{x}(\cosh\xi+\sinh\xi)=\bar{x}\exp(\xi)
\end{equation}
Following similar procedures for all other cases, the solutions for
$\{x,z\}$ and $\{y,w\}$ can be expressed as
\begin{equation}
	\{x,z\}=\begin{cases}
		\textrm{sign}(\bar{x})\{\sqrt{a}\cosh(\xi-\bar{\varphi}_{x}),-\sqrt{c}\sinh(\xi-\bar{\varphi}_{x})\} & \textrm{if }a>0;\\
		\textrm{sign}(\bar{z})\{-\sqrt{-a}\sinh(\xi-\bar{\varphi}_{z}),\sqrt{-c}\cosh(\xi-\bar{\varphi}_{z})\} & \textrm{if }a<0\\
		\bar{x}\exp(\xi)\{1,-1\} & \textrm{if }a=0\;\textrm{and}\;b\neq0
	\end{cases}\label{eq:xz-sol-g}
\end{equation}
and 
\begin{equation}
	\{y,w\}=\begin{cases}
		\textrm{sign}(\bar{y})\{\sqrt{b}\cosh(\xi+\bar{\varphi}_{y}),\sqrt{d}\sinh(\xi+\bar{\varphi}_{y})\} & \textrm{if }b>0;\\
		\textrm{sign}(\bar{w})\{\sqrt{-b}\sinh(\xi+\bar{\varphi}_{w}),\sqrt{-d}\cosh(\xi+\bar{\varphi}_{w})\} & \textrm{if }b<0\\
		\bar{y}\exp(-\xi)\{1,-1\} & \textrm{if }b=0\;\textrm{and}\;a\neq0
	\end{cases} \label{eq:yw-sol-g}
\end{equation}
where
\[
\tanh\bar{\varphi}_{x}=\dfrac{1}{p}\dfrac{\bar{z}}{\bar{x}}=1/\tanh\bar{\varphi}_{z},\;\textrm{and}\;\tanh\bar{\varphi}_{y}=\dfrac{1}{q}\dfrac{\bar{w}}{\bar{y}}=1/\tanh\bar{\varphi}_{w}.
\]
For general vertices that are non-flat-foldable and have no collinear creases (i.e., $a\neq0$ and $b\neq0$), by substituting the initial state in Eq. (\ref{eq:initial-state}) into Eqs. (\ref{eq:xz-sol-g})
and (\ref{eq:yw-sol-g}), we obtain
\begin{equation}
	\{x,z\}=\begin{cases}
		\varepsilon\{\sqrt{a}\cosh\xi,-\sqrt{c}\sinh\xi\} & \textrm{if }a>0,\\
		\varepsilon\{-\sqrt{-a}\sinh\xi,\sqrt{-c}\cosh\xi\} & \textrm{if }a<0
	\end{cases}\label{eq:xz-sol-cosh}
\end{equation}
and
\begin{equation}
	\{y,w\}=\begin{cases}
		\textrm{sign}(\bar{y})\{\sqrt{b}\cosh(\xi-\varphi),\sqrt{d}\sinh(\xi-\varphi)\} & \textrm{if }b>0,\\
		-\textrm{sign}(\bar{y})\{\sqrt{-b}\sinh(\xi-\varphi),\sqrt{-d}\cosh(\xi-\varphi)\} & \textrm{if }b<0
	\end{cases}\label{eq:yw-sol-cosh}
\end{equation}
where 
\[
\textrm{sign}(\bar{y})=\varepsilon\textrm{sign}(\sin(\gamma+\delta))\quad\textrm{and}\quad\varphi=\textrm{arctanh}(h^{-\textrm{sign}(ab)}).
\]
Recalling that $\cosh x=\dfrac{\exp(x)+\exp(-x)}{2}$ and $\sinh x=\dfrac{\exp(x)-\exp(-x)}{2}$,
Eqs. (\ref{eq:xz-sol-cosh}) and (\ref{eq:yw-sol-cosh}) can be rewritten as in Eq. (\ref{eq:xzyw-sol-exp}).

For flat-foldable vertices, i.e., $\alpha+\gamma=\pi$, Eqs. (\ref{eq:eop}) and (\ref{eq:ead}) lead to
\begin{equation}
	z=-\lambda x\quad\textrm{and}\quad w=\lambda y=\dfrac{\sin\alpha-\lambda\sin\beta}{\sin(\alpha+\beta)}x\label{eq:flatf-linear}
\end{equation}
where $\lambda=\pm$. Hence, all four angles are linearly related.
In addition, there is $p=q=1$ for flat-foldable vertices. 
The parametric solution obtained from Eqs. (\ref{eq:LT-r13}) and (\ref{eq:curve-r24}) is
\begin{equation}
	\{x,z,y,w\}=\bar{x}\{1,-\lambda1,\dfrac{\lambda\sin\alpha-\sin\beta}{\sin(\alpha+\beta)},\dfrac{\sin\alpha-\lambda\sin\beta}{\sin(\alpha+\beta)}\}\exp(\lambda\xi).\label{eq:sol-flatf}
\end{equation}
In Eq. (\ref{eq:sol-flatf}), $\lambda=+$ denotes mode-1, while $\lambda=-$ denotes mode-2.

\subsection*{Examples: kinematics of typical collinear and flat-foldable vertices}

\begin{figure}
	\centering
	\includegraphics[width=0.9\textwidth]{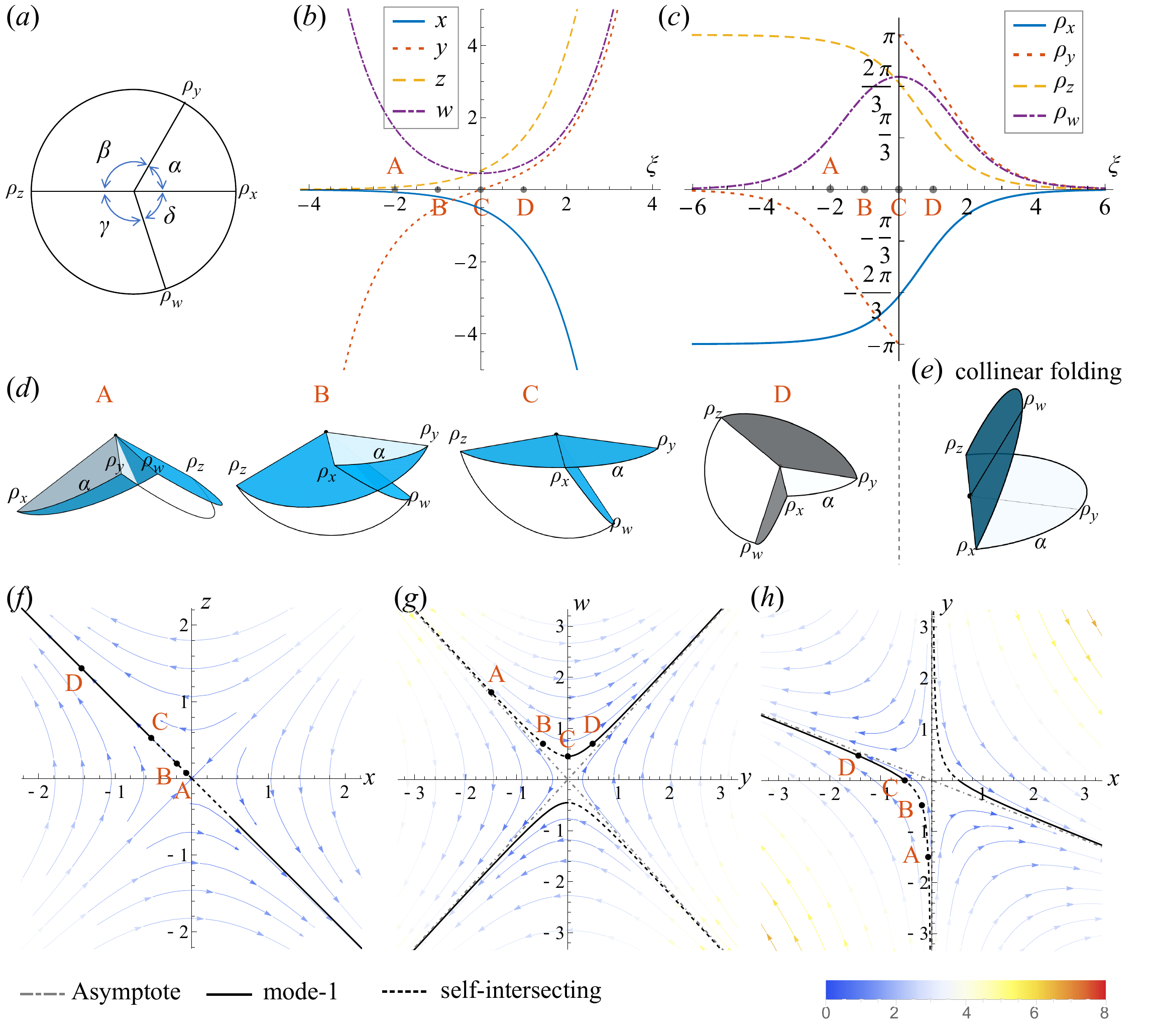}
	\caption{\label{fig:collinear-vertex}
		Kinematics of a collinear vertex with sector angles $\{\alpha,\beta,\gamma,\delta\}=\{\pi/3,2\pi/3,3\pi/5,2\pi/5\}$. 
		(a) shows the crease pattern where crease-$x$ and -$z$ are collinear.
		(b) and (c) plot the fold angles in terms of $\cot(\rho_{r}/2)$ and
		$\rho_{r}$ versus $\xi$, respectively, for $r=x,y,z$, and $w$.
		(d) shows the folded forms at $\xi=\{-2,-1,0,1\}$. (e) is the trivial
		folding along the collinear creases. (f), (g), and (h) show the kinematic path and vector field on the $x$-$z$, $y$-$w$, and $x$-$y$ planes, respectively.}
\end{figure}
\begin{figure}
	\centering
	\includegraphics[width=0.9\textwidth]{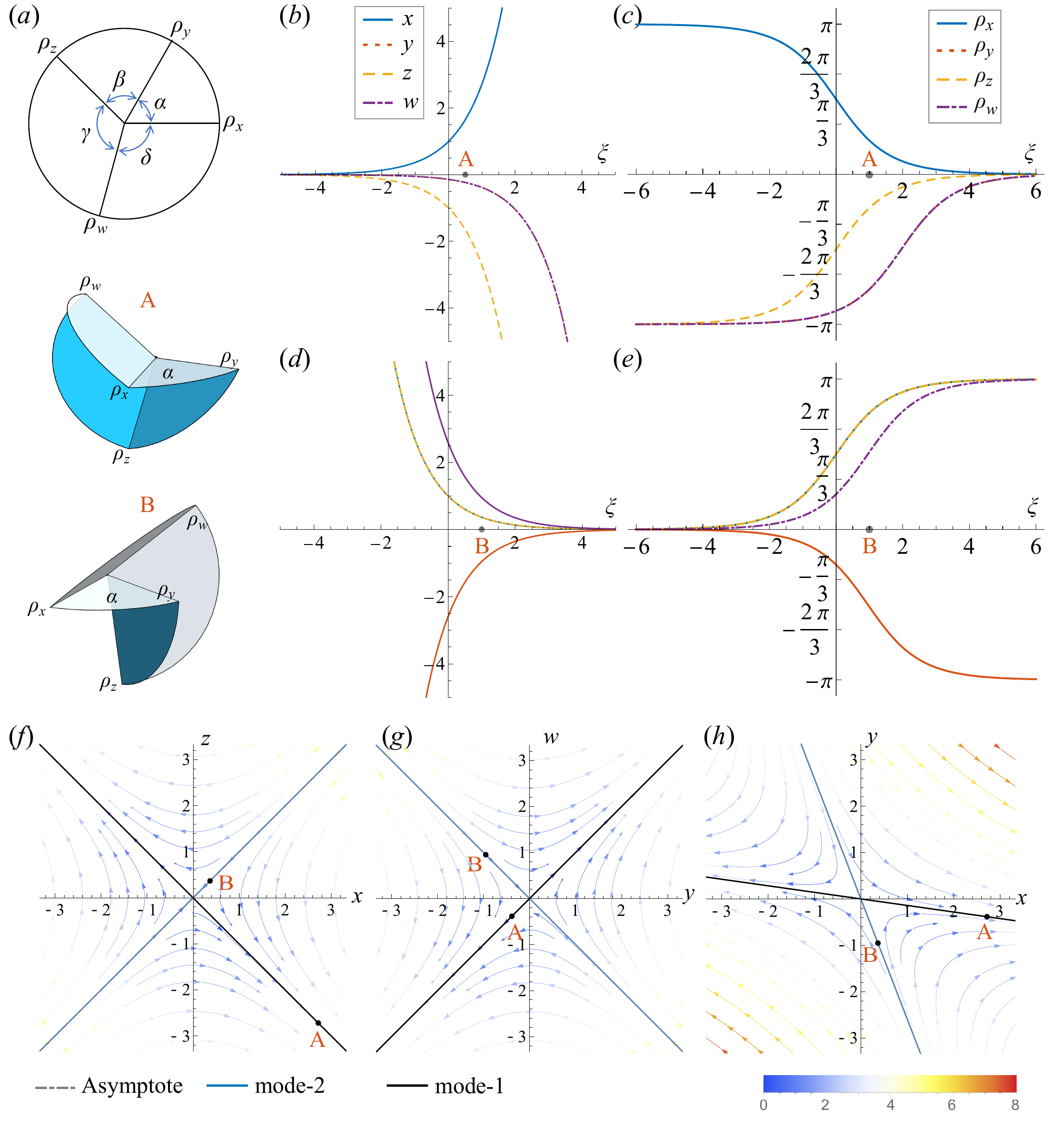}
	\caption{\label{fig:flat-foldable-vertex}Kinematics of a flat-foldable vertex with sector angles $\{\alpha,\beta,\gamma,\delta\}=\{\pi/3,5\pi/12,2\pi/3,7\pi/12\}$.
		(a) shows the crease pattern (top), folded forms of mode-1 (middle)
		and mode-2 (bottom). (b) and (c) depict the fold angles of mode-1
		in terms of $\cot(\rho_{r}/2)$ and $\rho_{r}$ versus $\xi$, respectively,
		for $r=x,y,z$, and $w$. (d) and (e) are the counterparts to (b)
		and (c) for mode-2. (f), (g), and (h) show the kinematic path and vector field on the
		$x$-$z$, $y$-$w$, and $x$-$y$ planes, respectively.}
\end{figure}
This section presents examples demonstrating the kinematics of collinear and flat-foldable vertices which are commonly utilized in origami design and applications. 
Fig. \ref{fig:collinear-vertex} presents
the kinematics of the collinear vertex with sector angles $\{\alpha,\beta,\gamma,\delta\}=\{\pi/3,2\pi/3,3\pi/5,2\pi/5\}$.
From Fig. \ref{fig:collinear-vertex}(b) and (c), fold angles $y$
and $w$ follow sinh and cosh functions as those of general vertices;
$\rho_{x}$($=-\rho_{z}$) increases from $-\pi$ to 0 as $\xi$ varies
from $-\infty$ to $+\infty$, indicating that collinear vertex folds
flat when $\xi\to+\infty$ but not when $\xi\to-\infty$. This fact
is also reflected in Fig. \ref{fig:collinear-vertex}(f), (g), and
(h), where the kinematic path consists solely of mode-1 and self-intersecting
configurations. Fig. \ref{fig:collinear-vertex}(f) clearly demonstrates
the degenerated linear relationship between the collinear fold angles
$x$ and $z$.

Fig. \ref{fig:flat-foldable-vertex} presents the kinematics of the
flat-foldable vertex with sector angles $\{\alpha,\beta,\gamma,\delta\}=\{\pi/3,5\pi/12,2\pi/3,7\pi/12\}$.
Figs.\ref{fig:flat-foldable-vertex}(b) and (c) plot the fold angles
for mode-1 in terms of $\cot(\rho_{r}/2)$ and $\rho_{r}$ versus
$\xi$, respectively, while (d) and (e) are the counterparts for mode-2.
Representative folded forms for mode-1 and -2 are depicted in the
middle and bottom of Fig. \ref{fig:flat-foldable-vertex}(a). Unlike
general vertices, mode-1 and mode-2 of a flat-foldable vertex are
disconnected, as the two modes cannot be interchanged when $\xi$
varies from $-\infty$ to $+\infty$; instead, the mode is specified
by $\sigma$ in Eq. (\ref{eq:sol-flatf}). Fig. \ref{fig:flat-foldable-vertex}(f),
(g), and (h) show that all pairs of fold angles, particularly adjacent
fold angles such as $x$ and $y$, exhibit a linear relationship.

\subsection*{Composition of tangent vectors}
\subsubsection*{The expression for $\mathbf{Y}_{1,n-1}$}
This section derives the expression for $\mathbf{Y}_{1,n-1}$ in
Eqs. (\ref{eq:dx1dt-dxndt}) and (\ref{eq:dx1dt-dxndt-entry}).
For vertex-1 and -2 as shown in Fig. (\ref{fig:polyhedron-N}), from
Eq. (\ref{eq:curve-r12}), there are
\begin{equation}
	\dfrac{dy_{1}}{d\xi_{1}}=\sigma_{1}\dfrac{dy_{1}}{dt}=v_{1}x_{1}-h_{1}y_{1}\;\;\textrm{and}\;\;\dfrac{dx_{2}}{d\xi_{2}}=\sigma_{2}\dfrac{dx_{2}}{dt}=h_{2}x_{2}+u_{2}y_{2}\label{eq:dy1dt-dx2dt}
\end{equation}
Since $x_{2}=y_{1}$ and $\dfrac{dx_{2}}{dt}=\dfrac{dy_{1}}{dt}$,
we have
\begin{equation}
	x_{2}=y_{1}=\dfrac{\sigma_{1}v_{1}x_{1}-\sigma_{2}u_{2}y_{2}}{\sigma_{1}h_{1}+\sigma_{2}h_{2}}\label{eq:sol-x2y1}
\end{equation}
Substituting Eq. (\ref{eq:sol-x2y1}) into Eq. (\ref{eq:LT-r12}), there is
\begin{equation}
	\begin{Bmatrix}\dfrac{dx_{1}}{dt}\\
		\dfrac{dy_{2}}{dt}
	\end{Bmatrix}=\dfrac{1}{\sigma_{1}h_{1}+\sigma_{2}h_{2}}\begin{bmatrix}1+\sigma_{1}h_{1}\sigma_{2}h_{2} & -\sigma_{1}u_{1}\sigma_{2}u_{2}\\
		\sigma_{1}v_{1}\sigma_{2}v_{2} & -1-\sigma_{1}h_{1}\sigma_{2}h_{2}
	\end{bmatrix}\begin{Bmatrix}x_{1}\\
		y_{2}
	\end{Bmatrix}
	=\begin{bmatrix}h_{1,2} & u_{1,2}\\v_{1,2} & -h_{1,2}\end{bmatrix} \begin{Bmatrix}x_{1}\\y_{2}\end{Bmatrix}
	=\mathbf{Y}_{1,2}\begin{Bmatrix}x_{1}\\y_{2}
	\end{Bmatrix}\label{eq:Xhc12}
\end{equation}
where $\mathbf{Y}_{1,2}$ is self-defined and its entries are denoted by $h_{1,2}, u_{1,2}$, and $v_{1,2}$. 
Besides, the phase shift $\varphi$ in Eq. (\ref{eq:xzyw-sol-exp}) can be rewritten as
\begin{equation}
	\sigma h=(\textrm{tanh}(\sigma\varphi))^{-\textrm{sign}(ab)}=\frac{\exp(\sigma\varphi)+\textrm{sign}(ab)\exp(-\sigma\varphi)}{\exp(\sigma\varphi)-\textrm{sign}(ab)\exp(-\sigma\varphi)}
	\label{eq:sigma_h}
\end{equation}
Substituting Eq. (\ref{eq:sigma_h}) into Eq. (\ref{eq:Xhc12}), there is
\begin{equation}
	\begin{aligned}
	h_{1,2}&
	=\dfrac{1+\sigma_{1}h_{1}\sigma_{2}h_{2}}{\sigma_{1}h_{1}+\sigma_{2}h_{2}}
	=\frac{1+(\textrm{tanh}(\sigma_{1}\varphi_{1}))^{-\textrm{sign}(a_{1}a_{2})}(\textrm{tanh}(\sigma_{2}\varphi_{2}))^{-\textrm{sign}(a_{2}a_{3})}}{(\textrm{tanh}(\sigma_{1}\varphi_{1}))^{-\textrm{sign}(a_{1}a_{2})}+(\textrm{tanh}(\sigma_{2}\varphi_{2}))^{-\textrm{sign}(a_{2}a_{3})}}\\
	&=(\textrm{tanh}(\sigma_{1}\varphi_{1}+\sigma_{2}\varphi_{2}))^{-\textrm{sign}(a_{1}a_{3})}
	\end{aligned}
\end{equation}
Thus, $\mathbf{Y}_{1,2}$ can be rewritten as
\begin{equation}
	\mathbf{Y}_{1,2}=\begin{bmatrix}h_{1,2} & u_{1,2}\\
		v_{1,2} & -h_{1,2}
	\end{bmatrix}=\begin{bmatrix}(\textrm{tanh}(\sigma_{1}\varphi_{1}+\sigma_{2}\varphi_{2}))^{-\textrm{sign}(a_{1}b_{2})} & \frac{-\sigma_{1}u_{1}\sigma_{2}u_{2}}{(\textrm{tanh}(\sigma_{1}\varphi_{1}))^{-\textrm{sign}(a_{1}a_{2})}+\sigma_{2}h_{2}}\\
		\frac{\sigma_{1}v_{1}\sigma_{2}v_{2}}{(\textrm{tanh}(\sigma_{1}\varphi_{1}))^{-\textrm{sign}(a_{1}a_{2})}+\sigma_{2}h_{2}} & -(\textrm{tanh}(\sigma_{1}\varphi_{1}+\sigma_{2}\varphi_{2}))^{-\textrm{sign}(a_{1}b_{2})}
	\end{bmatrix}.\label{eq:Y12}
\end{equation}
The composition process from Eq. (\ref{eq:dy1dt-dx2dt}) to Eq. (\ref{eq:Xhc12})
can be repeated untill vertex-($n-1$), resulting in Eq. (\ref{eq:dx1dt-dxndt}).

\subsubsection*{Proof of Eq. (\ref{eq:tangent-vector-compat})}
We first consider the (1,1)-entry of $\mathbf{Y}_{1,n-1}$. 
With Eq. (\ref{eq:phase-shift-0}) and $b_{n}=a_{1}$, it follows that
\[
h_{1,n-1}
=[\textrm{tanh}({\sum_{i=1}^{n-1}}\sigma_{i}\varphi_{i})]^{-\textrm{sign}(a_{1}a_{n})}
=-[\textrm{tanh}(\sigma_{n}\varphi_{n})]^{-\textrm{sign}(b_{n}a_{n})}
=-\sigma_{n}h_{n}.
\]
Then, we consider the (1,2)-entry of $\mathbf{Y}_{1,n-1}$. 
With $u$ defined in Eq. (\ref{eq:LT-r12}), Eq. (\ref{eq:dydx}) can be rewritten as
\begin{equation}
	\dfrac{dy}{dx}=\dfrac{(\tanh\xi)^{-\textrm{sign}(a)}-h}{u}=\dfrac{[\tanh(\sigma\xi)]^{-\textrm{sign}(a)}-\sigma h}{\sigma u}.\label{eq:dydx-u}
\end{equation}
From Eq. (\ref{eq:dydx-u}), we have
\begin{equation}
	\dfrac{dy_{1}}{dx_{1}}\dfrac{dy_{2}}{dx_{2}}
	\cdots\dfrac{dy_{n-1}}{dx_{n-1}}
	=\frac{(\tanh t)^{-\textrm{sign}(a_{1})}-\sigma_{1}h_{1}}{\sigma_{1}u_{1}}{\prod_{i=2}^{n-1}}
	\frac{[\tanh({\sum_{j=1}^{i-1}}\sigma_{j}\varphi_{j}-t)]^{-\textrm{sign}(a_{i})}+\sigma_{i}h_{i}}
	{-\sigma_{i}u_{i}}.\label{eq:Y-e12-a}
\end{equation}
Besides, similar to the derivation of Eq. (\ref{eq:dydx}), it can
be derived from Eqs. (\ref{eq:ead_1}), (\ref{eq:yw-sol-cosh}), and
(\ref{eq:curve-r12}) that
\begin{equation}
	\dfrac{dy}{dx}=\dfrac{\sin(\beta+\gamma)}{\sin\delta}\left[\dfrac{1}{h}(\tanh(\xi-\varphi))^{-\textrm{sign}(b)}+1\right]^{-1}=\dfrac{v}{[\tanh(\xi-\varphi)]^{-\textrm{sign}(b)}+h}.\label{eq:dydx-v}
\end{equation}
Equations (\ref{eq:dydx-v}) and (\ref{eq:phase-shift-0}) yield
\begin{equation}
	\dfrac{dy_{n}}{dx_{n}}=\frac{v_{n}}{[\tanh(\sigma_{n}(t-\sum_{j=1}^{n-1}\sigma_{j}\varphi_{j})-\varphi_{n})]^{-\textrm{sign}(a_{1})}+h_{n}}=\frac{\sigma_{n}v_{n}}{(\tanh t)^{-\textrm{sign}(a_{1})}+\sigma_{n}h_{n}}.\label{eq:Y-e12-b}
\end{equation}
Thus,
\begin{equation}
	\lim_{t\to0}\dfrac{dy_{1}...dy_{n-1}}{dx_{1}...dx_{n-1}}\dfrac{dy_{n}}{dx_{n}}=\frac{\sigma_{n}v_{n}}{\sigma_{1}u_{1}}\lim_{t\to0}{\prod_{i=2}^{n-1}}\frac{[\tanh({\sum_{j=1}^{i-1}}\sigma_{j}\varphi_{j}-t)]{}^{-\textrm{sign}(a_{i})}+\sigma_{i}h_{i}}{-\sigma_{i}u_{i}}\frac{(\tanh t)^{-\textrm{sign}(a_{1})}-\sigma_{1}h_{1}}{(\tanh t)^{-\textrm{sign}(a_{1})}+\sigma_{n}h_{n}}.\label{eq:Y-e12}
\end{equation}
If $a_{1}>0$, Eq(\ref{eq:Y-e12}) is
\[
\lim_{t\to0}\dfrac{dy_{1}...dy_{n-1}}{dx_{1}...dx_{n-1}}\dfrac{dy_{n}}{dx_{n}}=
\frac{\sigma_{n}v_{n}}{\sigma_{1}u_{1}}{\prod_{i=2}^{n-1}}\frac{[\tanh({\sum_{j=1}^{i-1}}\sigma_{j}\varphi_{j})]{}^{-\textrm{sign}(a_{1}a_{i})}+\sigma_{i}h_{i}}{-\sigma_{i}u_{i}}=\frac{\sigma_{n}v_{n}}{u_{1,n-1}}
\]
Thus $u_{1,n-1}=\sigma_{n}v_{n}$.

If $a_{1}<0$, there is
\begin{equation}
	(\textrm{tanh}({\sum_{j=1}^{i-1}}\sigma_{j}\varphi_{j}-t))^{^{-\textrm{sign}(a_{i})}}+\sigma_{i}h_{i}=
	\left[(\textrm{tanh}({\sum_{j=1}^{i-1}}\sigma_{j}\varphi_{j}-t))^{^{-\textrm{sign}(a_{1}a_{i})}}+\sigma_{i}h_{i}\right]f_{i}
\end{equation}
with
\[
f_{i}=\frac{(\textrm{tanh}({\sum_{j=1}^{i}}\sigma_{j}\varphi_{j}-t))^{^{-\textrm{sign}(a_{1}a_{i+1})}}}{(\textrm{tanh}({\sum_{j=1}^{i-1}}\sigma_{j}\varphi_{j}-t))^{^{-\textrm{sign}(a_{1}a_{i})}}}.
\]
Noting that
\[
\lim_{t\to0}f_{2}...f_{n-1}=\lim_{t\to0}\frac{[\textrm{tanh}({\sum_{j=1}^{n-1}}\sigma_{j}\varphi_{j}-t)]^{-\textrm{sign}(a_{1}a_{n})}}{[\textrm{tanh}(\sigma_{1}\varphi_{1}-t)]^{-\textrm{sign}(a_{1}a_{2})}}=
\frac{-\sigma_{n}h_{n}}{\sigma_{1}h_{1}}
\]
Eq(\ref{eq:Y-e12}) is
\begin{equation}
	\lim_{t\to0}\dfrac{dy_{1}...dy_{n-1}}{dx_{1}...dx_{n-1}}\dfrac{dy_{n}}{dx_{n}}=
	\frac{\sigma_{n}v_{n}}{u_{1,n-1}}\lim_{t\to0}f_{2}...f_{n-1}
	\frac{(\tanh t)^{-\textrm{sign}(a_{1})}-\sigma_{1}h_{1}}{(\tanh t)^{-\textrm{sign}(a_{1})}+\sigma_{n}h_{n}}=
	\frac{\sigma_{n}v_{n}}{u_{1,n-1}}
\end{equation}
Thus $u_{1,n-1}=\sigma_{n}v_{n}$. 
The (2,1)-entry of $\mathbf{Y}_{1,n-1}$ can be approached in a similar manner, or by using the properties that $\det(\mathbf{Y}_{1,n-1})=-1$ and $\det(\mathbf{X}_{h})=-1$. 
Hence, Eq. (\ref{eq:tangent-vector-compat}) is proved.

\subsection*{Construction of rigid-foldable quadrilateral of equimodular type}

This section provides an optimization scheme to construct the rigid-foldable
quadrilateral of equimodular type. 
For the case in Fig. \ref{fig:4gon}, the folding mode of the four vertices are chosen as $\{\sigma_{1},\sigma_{2},\sigma_{3},\sigma_{4}\}=\{+,-,+,-\}$,
i.e., the vertices are of mode $\{1,2,1,2\}$ when $t\to+\infty$.
The sector angles $\alpha_{1},\alpha_{2},\alpha_{3},\beta_{1}$, and
$\beta_{2}$ are set as the input parameters while $\gamma_{1},\gamma_{2},\gamma_{3},\gamma_{4},\beta_{3}$,
and $\beta_{4}$ are the unknown variables. 
The optimization scheme is
\begin{subequations}
	\begin{equation}
		\min\;(\varphi_{1}-\varphi_{2}+\varphi_{3}-\varphi_{4})^{2}+(\tau_{1}^{+}\tau_{2}^{-}\tau_{3}^{+}\tau_{4}^{-})^{2}\label{eq:construct-quad}
	\end{equation}
	\begin{equation}
		\textrm{subject to}\quad p_{i+1}=q_{i},\label{eq:cons-pq}
	\end{equation}
	\begin{equation}
		0<2\pi-\alpha_{i}-\beta_{i}-\gamma_{i}<\pi,\label{eq:cons-delta}
	\end{equation}
	\begin{equation}
		\textrm{and}\quad0<\gamma_{i}<\pi\quad\textrm{for}\quad i=1,2,3,4;\label{eq:cons-gamma}
	\end{equation}
	\begin{equation}
		\textrm{and}\quad0<\beta_{3}<\pi, \; 0<\beta_{4}<\pi.\label{eq:cons-beta}
	\end{equation}
\end{subequations}
Equations (\ref{eq:phase-shift-0}) and (\ref{eq:fam-gv-tau}) are
enforced through the objective function while Eq. (\ref{eq:equi-mag})
is included as the constraints. 
The constraints in Eqs. (\ref{eq:cons-delta}), (\ref{eq:cons-gamma}), and (\ref{eq:cons-beta}) are on the sector
angles for rigid-foldability. 
In the present case, we set $\{\alpha_{1},\alpha_{2},\alpha_{3},\beta_{1},\beta_{2}\}=\{1.3,1.5,1.7,1.6,\pi/3\}$,
and the optimization problem is solved using the interior point method.

\begin{figure}
	\centering
	\includegraphics[width=0.9\textwidth]{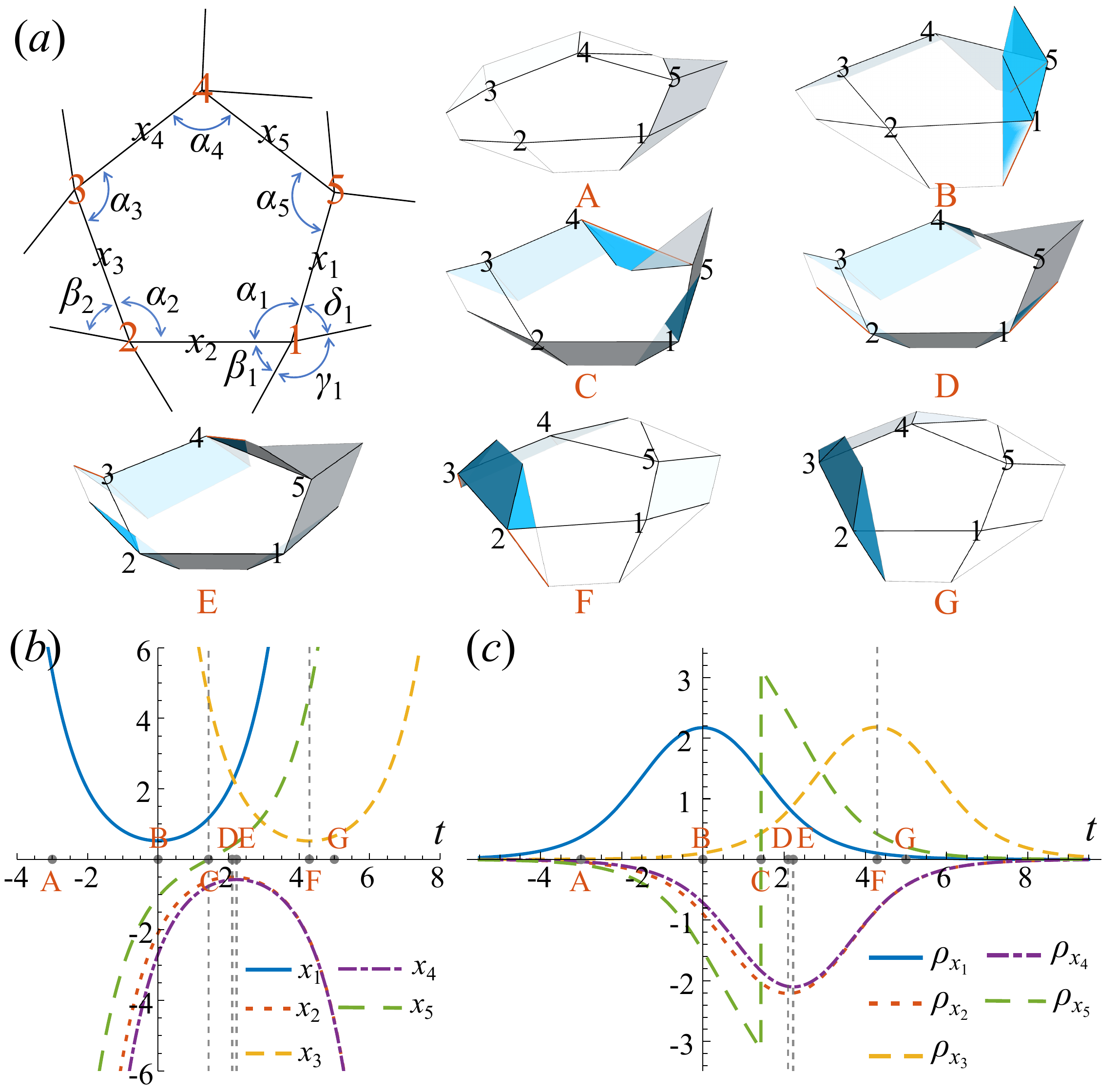}
	\caption{\label{fig:5gon}Rigid-foldable pentagon of equimodular type: fold
		angles and folded forms. In (a), the top-left subplot shows the crease
		pattern where the sector angles are given in Table.\ref{tab:sector-angles-k5},
		the remaining subplots labeled A through G are folded forms at $t=\{-3,0,\varphi_{5},\varphi_{1},\varphi_{1}+\varphi_{2}-\varphi_{3},\varphi_{1}+\varphi_{2},5\}$
		with $\{\varphi_{1},\varphi_{2},\varphi_{3},\varphi_{5}\}\approx\{2.100,2.190,2.064,1.432\}$.
		(b) plots the fold angles at the five creases of the central pentagon,
		i.e., $x_{i}$ for $i=1,2,3,4$ and 5, and (c) is the counterpart
		for $\rho_{x_{i}}$. Folded forms A through G in (a) correspond to
		those in (b) and (c). For the folded forms between B and F, at least
		one of the five vertices self-intersects. The folded forms B through
		F have binding folds, which are colored orange in (a) and located
		at crease-($z_{1},w_{5}$), crease-$x_{5}$, crease-($w_{1},z_{2}$),
		crease-($w_{3},z_{4}$), and crease-($w_{2},z_{3}$), respectively.}
\end{figure}

\begin{figure}
	\centering
	\includegraphics[width=0.9\textwidth]{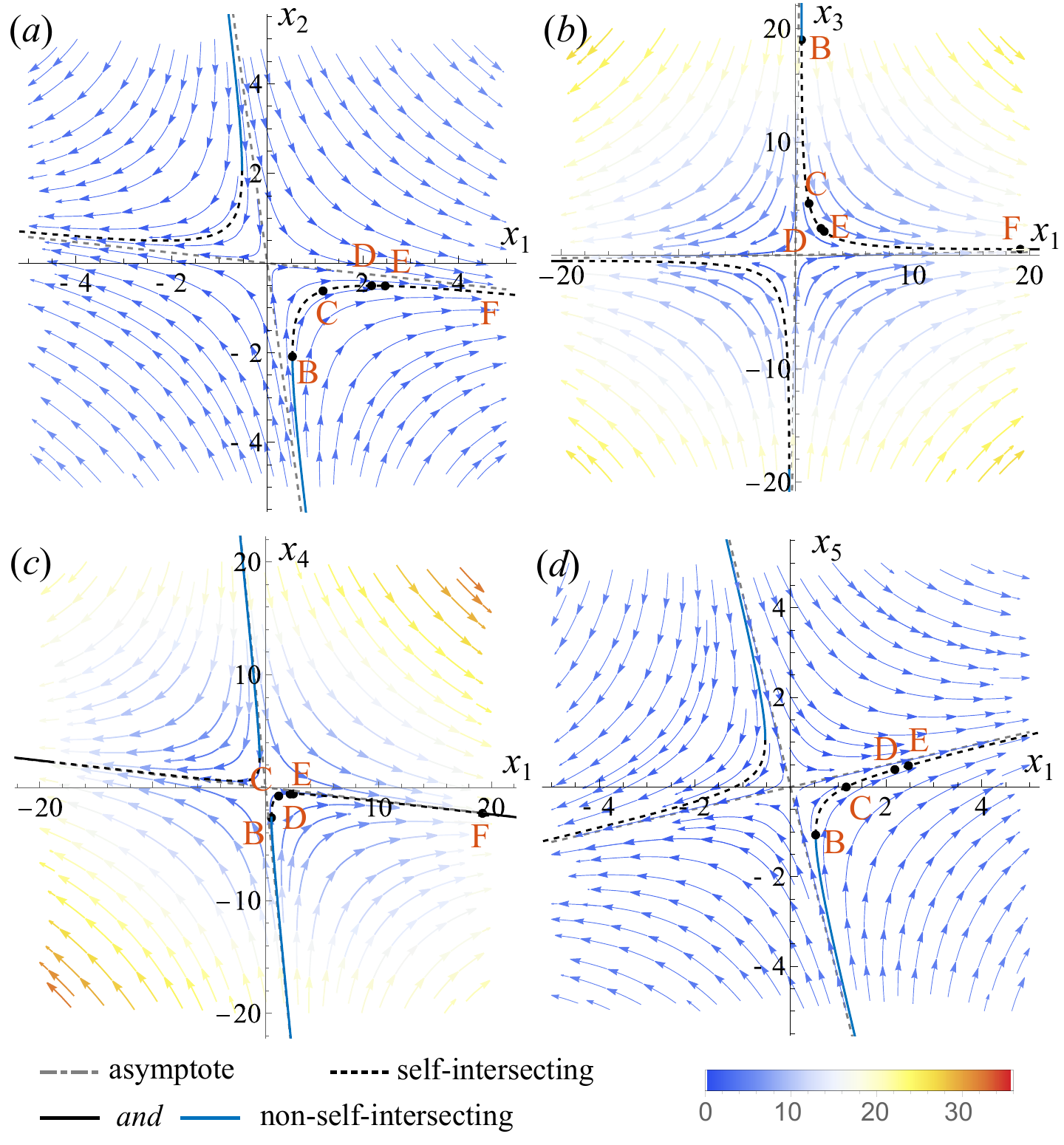}
	\caption{\label{fig:5gon-field}Rigid-foldable pentagon of equimodular type:
		kinematic path and vector field. (a), (b), (c), and (d) plot the results on the $x_{1}$-$x_{2}$, $x_{1}$-$x_{3}$, $x_{1}$-$x_{4}$, and $x_{1}$-$x_{5}$ planes, respectively.}
\end{figure}

The algorithm can also be adapted to construct rigid-foldable $n$-gon
of equimodular type. 
Table \ref{tab:sector-angles-k5}, Figs.\ref{fig:5gon}, and \ref{fig:5gon-field} are the results for a rigid-foldable pentagon, in which the folding mode of the
five vertices are chosen as $\{\sigma_{1},\sigma_{2},\sigma_{3},\sigma_{4},\sigma_{5}\}=\{+,+,-,-,-\}$.
The input parameters are $\alpha_{1},\alpha_{2},\alpha_{3},\alpha_{4},\beta_{1},\beta_{2},$
and $\beta_{3}$ (see Table \ref{tab:sector-angles-k5} for their
values) while the unknown variables are $\gamma_{1},\gamma_{2},\gamma_{3},\gamma_{4},\gamma_{5},\beta_{4}$,
and $\beta_{5}$.
\begin{table}
	\centering
	\caption{\label{tab:sector-angles-k5}Sector angles for the rigid-foldable pentagon of equimodular type}
	\begin{tabular}{|c|c|c|c|c|}
		\hline 
		Vertex & $\alpha$ & $\beta$ & $\gamma$ & $\delta$\tabularnewline
		\hline 
		\hline 
		1 & $53\pi/90$ & $17\pi/50$ & 2.278793368290275 & 1.086201429554792\tabularnewline
		\hline 
		2 & $11\pi/18$ & $33\pi/100$ & 2.301687189755826 & 1.024910364545366\tabularnewline
		\hline 
		3 & $3\pi/5$ & $101\pi/300$ & 2.320530515979933 & 1.020029672337214\tabularnewline
		\hline 
		4 & $29\pi/50$ & 0.5878598233529577 & 1.595663326954863 & 2.277538417789685\tabularnewline
		\hline 
		5 & $31\pi/50$ & 1.733163151547297 & 1.778269867159127 & 0.8239648432474907\tabularnewline
		\hline 
	\end{tabular}
\end{table}

\end{document}